# Evaluation of a multimode receiver with a photonic integrated combiner for satellite to ground optical communications

Vincent Billault, Jérôme Bourderionnet, Luc Leviandier, Patrick Feneyrou, Anaëlle Maho, Michel Sotom, Arnaud Brignon

*Abstract*— Multimode receivers based on spatial or modal diversity are promising architectures to mitigate in real time the atmospheric turbulence effects for free space optical (FSO) communications. In this paper, we evaluate and comment on the dynamical communication performances of a FSO mode diversity receiver, based on a spatial demultiplexer and a silicon photonic coherent combiner, for an optical link from a GEO satellite to an optical ground station (OGS). We simulate time series of distorted wavefronts received by the OGS and we show numerically that the coherent combination of spatial modes mitigate the signal fading compared to a conventional single mode fiber (SMF) receiver. We verify this property in a laboratory environment by generating the wavefronts corresponding to the use case with an atmospheric propagation channel emulator. Then we modulate the optical carrier prior to the wavefront emulator with 10G OOK and DPSK data sequences to measure the BER performance of the proposed receiver during the time series emulation. Finally, we study and comment on the influence of the number of modes combined and the wavelength multiplexing on the BER performances. We prove that the mode diversity receiver provide a higher collection efficiency, has better BER performances and much less synchronization losses.

## I. Introduction

Free space optical communications offer strong assets compared to RF links : a 1000 time higher available bandwidth with unlicensed spectrum (thanks to highly directive antennas) and a synergy with the terrestrial fibered network. However, for satellite to ground communication some important challenges remains to enable very high data rate free space optical (FSO) solutions [1, 2, 3]. Mainly, atmospheric turbulence in the light free space propagation induce beam wandering and scintillation that distort the optical wavefronts and degrade the beam optical coherence in the receiver antenna [4]. These phenomena cause signal fading at conventional single mode fiber (SMF) receivers output and limits the link communication performances. In the literature, two different solutions are commonly considered to reduce these effects: adaptive optics [5, 6, 7] and modal/spatial diversity receivers [8, 9, 10, 11, 12]. In the first approach, deformable mirrors shapes the incoming wavefront at the reception stage to mitigate the atmospheric effects. However, the complexity of the control algorithm limits in practice the correction loop bandwidth, and the correction can be inadequate for complex wavefront [7]. With spatial or modal diversity, the collection area of the communication receiver is increased with a multimode receiver or multiple SMF receivers. The communication signal modulates each optical carrier collected from the different modes/apertures but with an intensity and phase that fluctuates with time because of the dynamical evolution of the atmospheric turbulence. To mitigate these fluctuations, the modulation signals are combined after [8, 9, 12] or before [10, 11] optical to electrical conversion. Multimode receivers decrease the signal fading that is unavoidable with SMF receivers, but the combination of the multitude of optical or electrical signals with components off the shelf is not easily scalable to a large number of signals and usually generates additional losses. Photonic integrated circuits (PICs) have been proposed for coherent combination [13, 14] to improve the performances of the combination part of the multimode receivers [15, 16]. Indeed, the integrated platforms allow to encompasses a large number of optic and electronic functions at mm scale with an outstanding mechanical and thermal stability, and with propagation losses on the chip including couplers and phase modulators as low as 0.5 dB/cm [17].

We presented in a previous work [18] a preliminary static evaluation of a multimode receiver collection efficiency with a PIC coherent combiner for FSO communications between a GEO satellite and an optical ground station (OGS). We simulated various phase and intensity wavefronts corresponding to the light spatial profile received in the OGS. Then we showed experimentally that the coupling efficiency of our proposed multimode receiver was statistically much less sensitive to atmospheric perturbation than a SMF receiver.

We present here the dynamical evaluation of the multimode communication receiver for a realistic FSO link issued from a GEO satellite. To this end, we generate time series of light propagation through the atmosphere from a GEO satellite to an OGS, and we compute the evolution of the spatial light

This work was partially supported by the VERTIGO project from the European Union's Horizon 2020 research and innovation program under grant agreement No. 822030. Vincent Billault, Jérôme Bourderionnet, Luc Leviandier, Patrick Feneyrou and Arnaud Brignon are with Thales Research and Technology France, 1 Avenue Augustin Fresnel, 91767 Palaiseau, France.

(e-mail: vincent.billault@thalesgroup.com). Anaëlle Maho and Michel Sotom are with Thales Alenia Space, 26 Avenue J.F. Champollion – B.P. 33787, 31037 Toulouse Cedex 1, France. This work has been submitted to the IEEE for possible publication. Copyright may be transferred without notice, after which this version may no longer be accessible.

distribution in the aperture of the OGS along time. We show in particular that the coupling efficiency depends a lot of the number of mode combined, and that coupling to a SMF FSO receiver induce strong localized fading events. We then use a wavefront emulator to reproduce in laboratory environment the FSO link, and we measure the evolution of the coupling efficiency with time for the proposed receiver as well as a SMF FSO receiver. We compare the dynamical communication performances of the two receivers with the measurement of the Bit Error Rate (BER) at the two receivers output with 10 Gb/s OOK and DPSK data sequences on the emulated FSO link under various disturbance condition. We confirm that the multimode receiver is much more robust to perturbation for both modulation formats and has less synchronization loss than the SMF receiver. Finally, we evaluate and comment on the influence of the number of modes combined and the wavelength multiplexing on the multimode receiver performances.

## II. GEO FSO LINK SIMULATION

### A. Spatial light distribution in the OGS aperture

A numerical model of atmosphere is used to emulate the propagation of the light from a GEO satellite to the OGS [18]. This model implements a beam propagation method with a series of alternating free space propagation and phase screens generated by Discrete Fourier Transform from a Kolmogorov - Von Karman – Tatarsky spectrum accounting for refraction indices fluctuations. Table 1 sums up the parameters used to simulate the evolution of the spatial light distribution in the aperture of the OGS along time. The field, whose wavelength is 1.55 µm, is computed in the receiver plan and projected on the fifteen Hermite-Gauss (HG) modes listed on Table 2. The size of the higher modes is chosen to fit the aperture diameter.

TABLE 1
INPUT PARAMETERS OF THE SIMULATION MODEL

| Turbulence strength | $8.7 \cdot 10^{-14}$ m$^{-2/3}$ |
|---|---|
| r0 | 7.7 cm |
| Wind speed | 47 m.s$^{-1}$ |
| Elevation of the line of sight | 30° |
| GEO (resp. OGS) telescope diameter | 40 cm (resp. 50 cm) |
| Field wavelength | 1.55 µm |

We extract from the time series the evolution of the energy distribution for the different HG modes in the collection aperture plan versus time. Fig. 1.a represent the relative energy distribution between the 15 first HG modes (see modes list in Table 2) for 100 successive frames. The atmospheric turbulence quickly redistribute the energy among the different HG modes over time (the corresponding frame rate is 1.5 kHz). However, on average the energy from the received wavefront couples more to the lower order spatial modes than the higher order (cf. Fig. 1.b): 32 % of the energy is coupled to the modes $HG_{00}$, $HG_{01}$ and $HG_{10}$. Table 2 shows the average relative power over the whole sequence duration of the time series (~1s) for the 15 first HG modes.

TABLE 2
AVERAGE RELATIVE POWER FOR THE HG MODES

| Modes HG$_{mn}$ | 00 | 01 | 10 | 02 | 11 | 20 | 03 | 12 |
|---|---|---|---|---|---|---|---|---|
| % relative power | 14.5 | 8.6 | 9.0 | 6.1 | 5.2 | 6.7 | 4.2 | 3.2 |
| Modes HG$_{mn}$ | 21 | 30 | 04 | 13 | 22 | 31 | 40 | Other |
| % relative power | 4.1 | 4.2 | 2.5 | 2.0 | 1.9 | 2.4 | 2.4 | 23 |

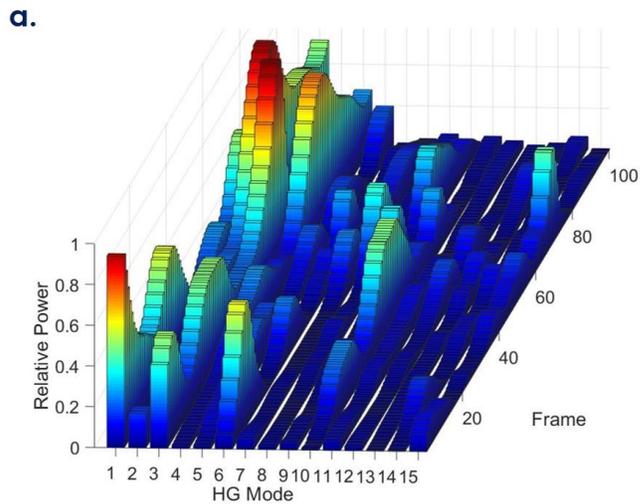

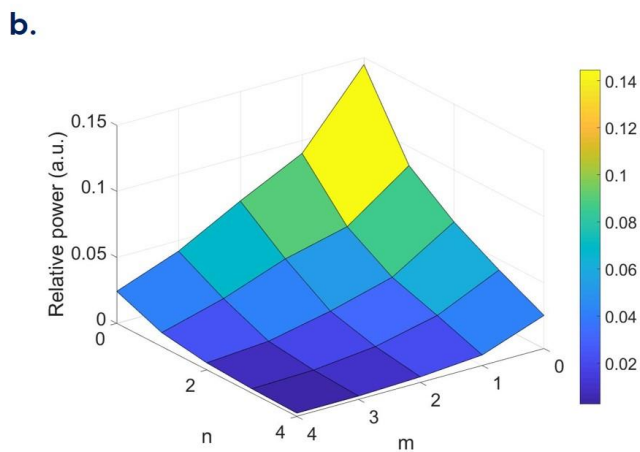

Fig. 1(a) Relative power distribution of the 15 first HG modes versus time for the GEO to OGS FSO link. (b) Relative power for the HG modes averaged on the time series duration.

### B. Coherent mode combining efficiency

With the spatial distribution of the received wavefronts in the OGS aperture, we compute the theoretical coupling efficiency (i.e. the ratio of the power at the receiver fiber output over the power at the receiver free space input) for a SMF and for a multimode (MM) receiver. We consider the SMF receiver optimized for the collection aperture (i.e : 81 % of coupling efficiency for an uniform waveform [20]). Concerning the MM receiver, to evaluate the influence of the number of modes coupled and combined, we compute the efficiency (Fig 2.a) for an increasing number of modes $HG_{mn}$ and for a lossless combination. As the energy distribution for the different spatial modes is not uniform, we increase the mode order (m + n) considered. The coupling efficiency of the SMF receiver output strongly varies along time compared to the MM receiver. The efficiency variation (difference between the maximum and the minimum) is > 30 dB for the whole time series sequence, which proves that signal fading with a simple SMF receiver is unavoidable. For the MM receiver, the average coupling efficiency and the efficiency variation depends on the number of mode considered (see Table 3). The higher the number of modes for combination are combined, the more stable is the coupling efficiency at the MM receiver output and the lower are the coupling losses. With 15 modes, the average coupling loss is 1.2 dB (76 % of the energy from the OGS aperture is collected) and the coupling efficiency variation is about 3 dB. The coherent combination of modes drastically reduce signal fading.

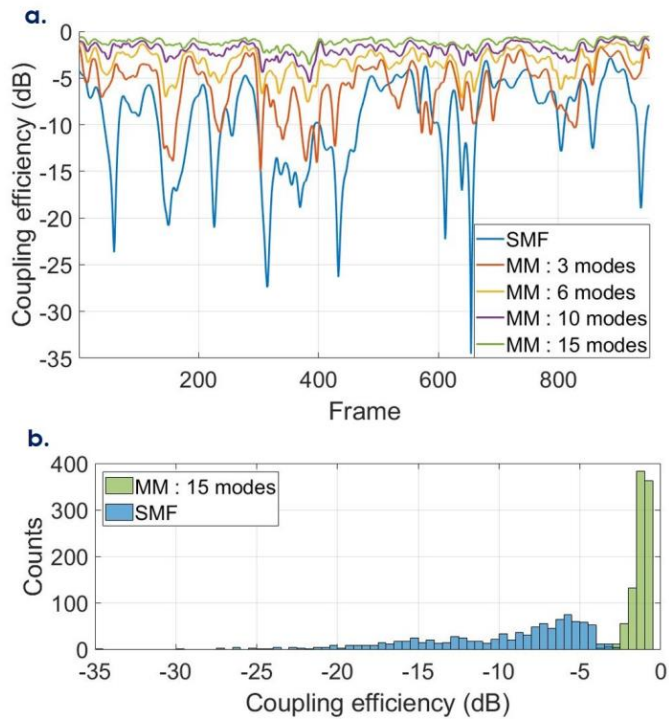

Fig. 2 Coupling efficiency for the SMF receiver (blue), and the MM receiver for different number of HGmn combined: $m + n < 2$ (red), $m + n < 3$ (yellow), $m + n < 4$ (purple), $m + n < 5$ (green). (a) Temporal evolution (b) histogram over the time serie.

TABLE 3
COUPLING LOSS AND MAX COUPLING EFFICIENCY VARIATION FOR A SMF A MM RECEIVER AVERAGED OVER THE TIME SERIES

| Number of modes | % of energy collected from the OGS aperture | Average coupling loss (dB) | Coupling efficiency variation (dB) |
| --- | --- | --- | --- |
| SMF | 17 % | -7.7 | 31.7 |
| MM 3 modes | 30 % | -5.3 | 13.4 |
| MM 6 modes | 49 % | -3.1 | 6.4 |
| MM 10 modes | 64 % | -1.9 | 4.8 |
| MM 15 modes | 76 % | -1.2 | 3.1 |

## III. EXPERIMENTAL SETUP

### A. Setup description

We developed an experimental setup to validate the use of a spatial demultiplexer and a coherent combiner for optical communications under the GEO link atmospheric turbulence conditions. In Table 3, we see that the average coupling efficiency increases as we increase the number of modes combined when considering a lossless combination. Following that principle, the optimal multimode receiver would be a receiver that can support the largest number of modes possible. However, the combination of several modes induce inherently some losses, so the gain in efficiency variation and average coupling loss as we increase the number of mode collected need to be compare to the extra losses that are due to the higher number of mode to be combined. Furthermore, as most of the energy is coupled to lower order spatial modes, the addition of higher order modes (in our case $m + n > 5$) has little effect on the coupling efficiency. The extra losses depend particularly on the multimode device that collects the light and the combination platform. We chose therefore a mutli-plane light conversion (MPLC) with 15 modes and a photonic integrated circuit to minimize the coupling loss of our multimode receiver. The multimode communication receiver and the wavefront emulator test bed to reproduce the atmospheric turbulence was proposed in [15].

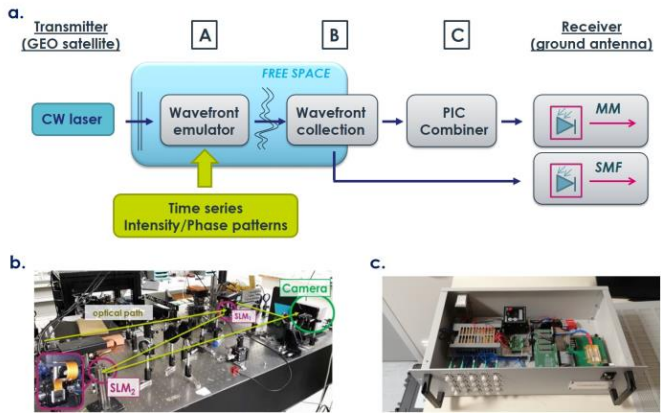

Fig. 3 (a) Experimental setup of the FSO multimode receiver the PIC combiner and the wavefront emulator test bed. (b) Free-space optical propagation channel emulator (c) picture of the photonic combining circuit with the control electronics.

The free space wavefront emulator (Fig. 3.a) generate arbitrary intensity and phase optical wavefronts, with the use of two nematic liquid crystal phase-only spatial light modulators (HOLOEYE LCOS SLMs). The phase and intensity pattern corresponding to the mode distribution of the light in the OGS antenna after propagation from the GEO satellite (cf. Fig. 2.a) are displayed onto the SLMs along time. As the display rate of the SLMs is below 5 Hz, we reduce the frame rate of the distorted wavefront to 3 Hz. The spatial demultiplexer (Fig. 3.b) is a MPLC (Cailabs) module that takes a FSO wavefront and returns its projection on a 15 HG modes basis (the 15 modes of Table 1), on 15 separated SMFs. The insertion losses of the MPLC, measured from the SMF inputs to the FSO output, is 1 dB in average.

Finally, the Silicon On Insulator (SOI) photonic integrated combiner (cf. Fig. 3C) combines the energy from the 15 SMFs with 5 cascaded Mach-Zehnder interferometers and carry the combined beam for either on-chip or outside the chip direct/balanced detection. The total PIC insertion loss at 1535 nm from any of the 15 fiber inputs to the on-chip direct or (I/Q) coherent detection is 7 dB. A Nelder Mead optimization algorithm on the PIC control electronics compensate for the intensity/phase fluctuations between the input SMFs. By generating a phase modulation on one PIC input with a phase modulator and correcting the phase modulation on chip with the Nelder mead algorithm, we measure that the algorithm corrects phase modulations up to 3 kHz for a two channels combination. This correction bandwidth is compliant for the dynamic compensation of atmospheric turbulence ($\leq$ kHz) [21].

## B. Dynamic coupling efficiencies

In a first experiment, we compare the light coupling efficiency of the MM and SMF receiver for the GEO to OGS FSO link use case (cf. Fig. 3). The transmitter injecting the wavefront emulator is a CW laser at 1556 nm followed by a high power optical amplifier (HPOA). For the SMF receiver, an afocal system adapts the size of the wavefront at the output of the wavefront emulator to a collimator to optimize the coupling efficiency. Without spatial modulation, the coupling loss of the SMF receiver is 4.6 dB while the coupling loss of the spatial demultiplexer and the PIC combiner is 8 dB (7 dB from the PIC, and 1 dB from the MPLC). Fig. 4 (a) shows the evolution of the power at the wavefront emulator output when it displays the successive wavefronts from the time series.

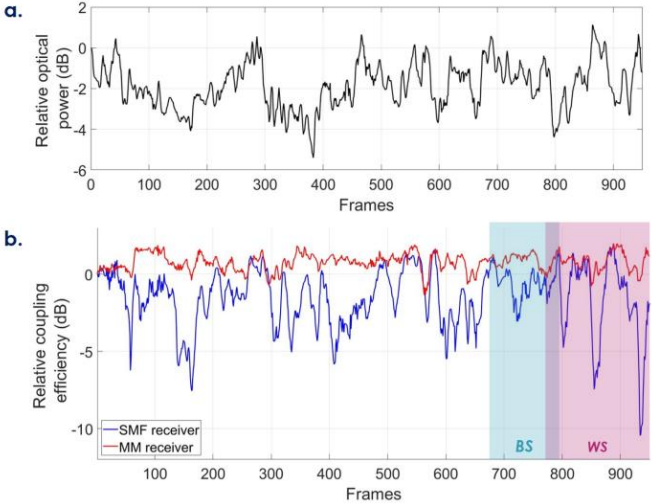

Fig. 4 (a) Optical power variation at the wavefront emulator output. (b) Relative coupling efficiency for the SMF (blue) and MM (red) receiver normalized to the coupling efficiency of the first frame. BS: best sequence WS: worst sequence.

Fig. 4.b shows the evolution of the relative coupling efficiency of the two receivers. The coupling of the SMF receiver strongly varies (12.7 dB max variation) as the coupling of the MM receiver remains almost constant (3.3 dB max variation) along the atmospheric turbulences profiles. This proves that the MM receiver is much more robust to intensity and phase variations for the satellite to ground communication. The experimental coupling efficiency of the MM receiver is close to the theoretical one (3.1 dB of max variation for 15 modes), however for the SMF receiver, the experimental coupling efficiency is much higher than the theoretical one (31.7 dB of max variation). This difference is likely due to the residual unmodulated light of the SLMs that is coupled to the SMF fiber.

## IV. MULTIMODE RECEIVER COMMUNICATION PERFORMANCES

### A. BER performances with the MM and SMF receiver for OOK and DPSK modulation

We then replace the CW laser with a communication transmitter (Tx) and add a demodulation module (Rx) at the output of the FSO receivers to compare the communications performances for the SMF and the MM receivers. The transmitter is composed of either one or several CW laser (to evaluate the influence of wavelength multiplexing, a Mach-Zehnder Modulator (MZM) and a HPOA (cf. Fig. 5). The demodulation module is either a photoreceiver for OOK modulated signals, or a delay line interferometer (DLI) with a dual-detector balanced receiver for DPSK demodulation. For both cases, the demodulation module includes also a low noise amplifier (LNOA) and an optical wavelength demultiplexer (WDM). The OOK and DPSK modulation are done with PRBS-15 data sequence, and the bit error rate (BER) is evaluated with a BER tester for different received optical power (ROP), i.e. the optical power received by the demodulation module, by adjusting the optical attenuation before the LNOA.

We compare the communication performances for three configurations (cf. Fig. 5):

- Back to back (BtB): the transmitter is followed by a variable optical attenuator (VOA) and then to the demodulation module.

- SMF: the transmitter is followed by the wavefront emulator, the SMF receiver, a VOA and then to the demodulation module.

- MM: the transmitter is followed by the wavefront emulator, the MPLC, the PIC combiner, a VOA and then to the demodulation module.

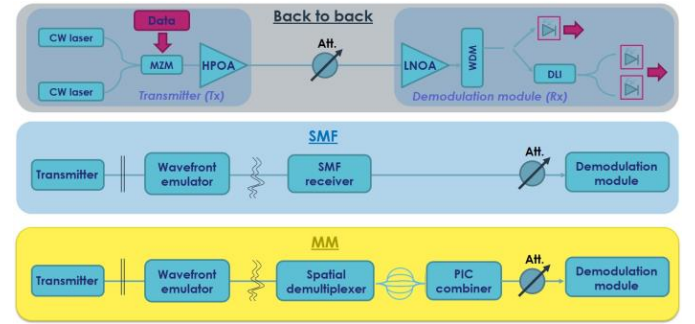

Fig. 5 Scheme of the communication performance measurement.

The BER as a function of the ROP is first evaluated with a static perturbation profile displayed on the SLM (Frame n°1 in the time series sequence) with OOK modulation with the same ROP in the three configurations (Fig. 6. a). The SMF receiver induce no penalty on the FSO link performance in that case. For the MM receiver, the link performance is identical to the BtB configuration at low ROP (< -38 dBm). However at high ROP (> -38 dBm), the BER reaches a floor. This originates from the combination algorithm that controls the coherent combiner: fast fluctuations of the phase/intensity between two channels induce 2 pi phase jumps in the command of the phase shifters [15]. These phase jumps generate residual power drop of the combined beam, which generate a constant error rate. The constant error rate is negligible at low received power, but become visible as the received power increases.

We then measure the cumulated BER with time series applied to the optical channel emulator. To investigate different disturbance condition, we identify two frame sequences (see Fig. 4) from the time series, and we measure the cumulated BER over the two sequences for the two receivers. In the first sequence (Frames 674 to 794 see Fig. 4.b, hereunder called best sequence) the coupling efficiency of the SMF fiber is relatively stable (max coupling efficiency variation ~ 3 dB). In the second sequence (Frames 770 to 951 see Fig. 4.b hereunder called worst sequence) the coupling efficiency of the SMF fiber varies a lot (max output power variation ~ 12 dB). As the ROP changes during the sequences for the two receivers, we change gradually the ROP for a static frame (Frame n°1 in the time series sequence) and then we measure the cumulated BER for the given attenuation. Fig. 6.b (resp Fig. 6.c) shows the resulting cumulated BER performances with OOK (resp. DPSK) modulation for the worst and best sequences and the static BER performance of the BtB link of Fig. 6. a.

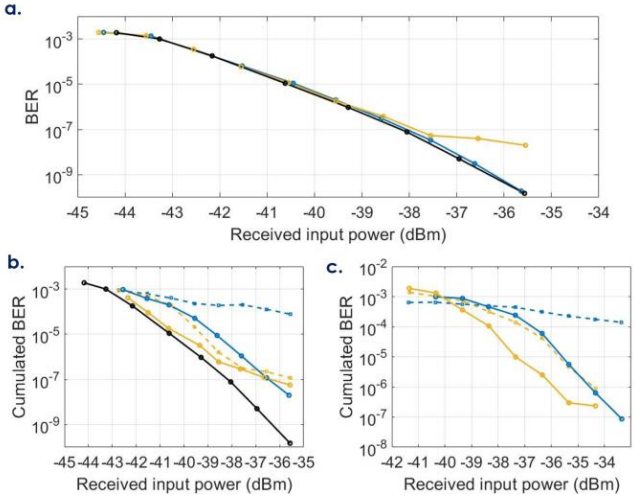

Fig. 6 (a) Cumulated BER as a function of the power at the Rx input for the frame n°1 with OOK modulation. black: BtB, blue: SMF, yellow: MM. Cumulated BER over the best (solid line) and worst (doted line) sequences at the Rx input for OOK (b) and DPSK (c) modulation. black: BtB, blue: SMF, yellow: MM

First for the OOK modulation, for the worst sequence, the cumulated BER of the SMF receiver has a floor at 1e-4 due to the low ROP during numerous frames of the sequence that induce a large number of error. On the contrary, for the MM receiver, the cumulated BER decreases as the ROP increases. The MM receiver induces in that case 1.6 dB of power penalty with respect to the BtB configuration at $10^{-4}$ BER. As in the static configuration, we see that the cumulated BER reaches a floor (~ 1e-7) at high received power. For the best sequence, as the coupling efficiency variation of the SMF receiver is much lower, the BER decreases as the ROP increases (the output power fluctuates less inducing less error during the sequence). Still, the penalty of the link is lower for the MM receiver ( 0.3 dB of power penalty at $10^{-4}$ BER) compare to the SMF receiver (1.8 dB of power penalty at $10^{-4}$ BER). This trend is reversed when the cumulated BER for the MM receiver reaches the BER floor. This two sequences show that the power is a key element of a BER perfomances, and that the MM receiver by reducing the received power fluctuations has better BER performances for different perturbation conditions. Also synchronization losses occur more often and last much longer for the SMF receiver compared to the MM receiver. Up to 13 seconds of interruptions per minute were measured against only a one-second interruption for the multimode receiver for the worst sequence. This trend is similar for the DPSK modulation:

the strong power fluctuations at the SMF receiver output induce a BER floor and for both sequences, the power penalty is much lower for the MM receiver than for the SMF receiver.

*B. Number of spatial modes combined*

The number of modes combined is critical for the design of the multimode receiver, as it affects both the absolute coupling efficiency and the coupling efficiency variation over a given perturbation. To measure the influence of the number of modes combined on the communication performances, we repeat the cumulated BER measurement for a different number of modes collected and routed to the coherent combiner. Fig. 7 shows the cumulated BER as a function of the ROP for the worst/best sequence and for OOK/DPSK modulation, for an increasing number of modes combined. The behavior of the cumulated BER versus ROP is similar for OOK and DPSK modulation format : for the best sequence, the power penalty compared to the BtB measurement gradually decreases as the number of modes increases (from 2.5 dB to 0.5 dB for a $10^{-5}$ cumulated BER for OOK modulation). For the worst sequence, the combination of six modes is not sufficient to avoid synchronization losses and a high BER floor, as the communication performance in that case is similar to the SMF receiver. However, starting from 10 collected and combined modes, the multimode receiver is resilient enough to reduce the power variation of the combined beam and decrease the cumulated BER penalty. The communication performances for 10 and 15 modes are almost identical, showing that the five additional higher order modes in that case are not improving significantly the performances. These experimental results concur with the numerical simulation: a high number of modes ($\geq 10$) should be considered to avoid signal fading and synchronization losses.

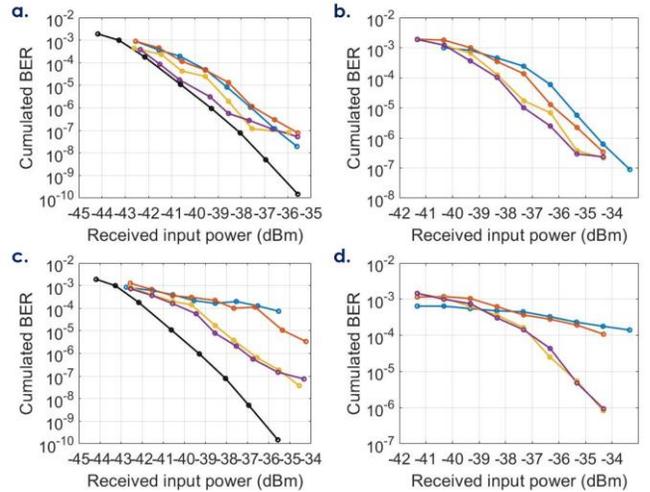

Fig. 7 Cumulated BER as a function of the power at the Rx input. blue: SMF, red: MM with 6 modes, yellow: MM with 10 modes, purple: MM with 15 modes, black: BtB (a) best (c) worst sequence for 10 Gb/s OOK signal. (b) best (d) worst sequence for 10 Gb/s DPSK signal.

*C. Wavelength multiplexing*

Finally, as optical feeder links are commonly considered with spectral multiplexing of a large number of channels [1] [2], we characterize the influence of the transmission of an additional wavelength on the same channel on the communication performances.

*1) Passive delay mismatch compensation*

On the receiver side, the simultaneous coherent combination of the 15 SMF fibers for different wavelength imposes a strict control of the delay mismatch between the channels. Considering an optical signal with a bandwidth $\Delta\nu$ injected in an optical phase array, the coherent combination is efficient as long as the delay mismatch $\tau$ obeys $|\Delta\nu\tau| \ll 1$ [22]. For two channels spaced by 100 GHz, the delay mismatch must be much lower than 3 mm. To match the delays between the 15 paths from the 15 SMF of the spatial demultiplexer, we add a variable optical delay line (VODL) between each SMF and the PIC inputs and we replace the transmitter with a C band broadband optical source with a tunable bandwidth. Then we combine the SMF fibers two by two and we scan the VODLs to find the position where the delays are matched (i.e. when the power at the combiner output is maximum) for different bandwidth of the broadband source. Fig. 8 illustrates the accuracy required on the path length differences for the PIC combiner for two paths as a function of the optical bandwidth at the input.

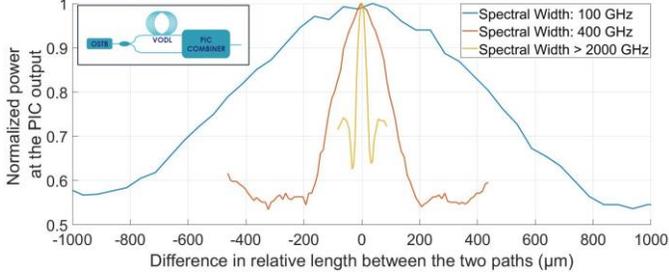

Fig. 8 Normalized power at the PIC combiner output for two paths versus the relative length difference between the two paths. Inset : Scheme of the delay mismatch correction setup (OSTB : optical source with a tunable bandwidth, VODL : variable optical delay line).

With the setting of the VODLs to the maximum optical bandwidth of Fig. 8, the coherent combiner is efficient for optical signals as wide as 16 nm around 1560 nm.

*2) BER performances of the GEO link for two wavelengths*

We use as transmitter two lasers spaced by 100 GHz and multiplexed in one single mode fiber (cf. Fig. 5). The power of the two lasers is the same and is equal half the power of the single laser in section IV A, to ensure that the transmitter with the 1 and 2 wavelengths configurations has the same combined output power. At the reception stage, the wavelengths are separated with a spectral demultiplexer and one modulated wavelength goes to detection stage. We perform the same BER measurement as in section IV A to see if the additional wavelength induce BER penalties on the communication performances.

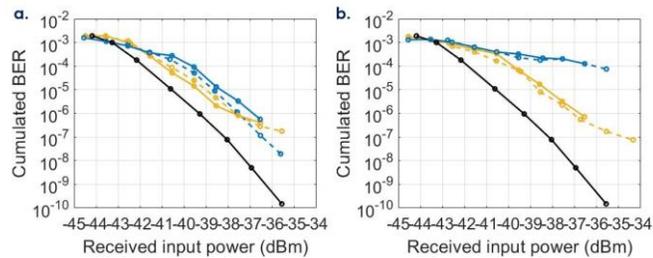

Fig. 9 Cumulated BER as a function of the power at the Rx input for (a) best (b) worst sequence for 10 Gb/s OOK modulation format with 1 (dotted lines) and 2 (solid line) wavelengths. blue: SMF, yellow: MM, black : BtB.

For the worst sequence, the influence of the additional wavelength is negligible. For the best sequence, a small difference arise for the SMF receiver at high-received power. Overall, for the different configurations tested, the used of the proposed multimode receiver for optical communication combined with spectral multiplexing on the same propagation channel has no additional penalty compared to a single wavelength.

*D. Atmospheric turbulences bandwidth discussion*

In our setup, the frame rate of the atmospheric perturbations was limited by the display rate of the SLM at 3 Hz. For the GEO satellite to an OGS use case, the frame rate that would correspond to the atmospheric turbulence is much higher. However, if we suppose the ROP constant over a time $\Delta t$, then the communication sequence can be written has a succession of N frames with a frame rate $F = 1/\Delta t$. And the cumulated bit error rate $\text{BER}_{\text{cum}}$ over the sequence writes:

$$\text{BER}_{\text{cum}} = \frac{F}{N} \cdot \sum_{Frame\ i=1}^{N} \frac{\text{BER}(P(i))}{F} = \frac{1}{N} \cdot \sum_{Frame\ i=1}^{N} \text{BER}(P(i)),$$

with P(i) the ROP at the frame i. In that case, the cumulated BER does not depends on the frame rate. This result is true as long as the ROP does not depends on $F$ which is valid for the SMF receiver, but not always valid for the MM receiver. Indeed, the power at the MM receiver output depends on the combination algorithm: the output power does not depend on the frame rate as long as the correction bandwidth $BW$ of the algorithm is larger than the frame rate.

## V. CONCLUSION

As a conclusion, the evaluation of a MM FSO receiver based on the coherent combination shows that it improves the FSO communication performances from a GEO station to an OGS compared to a standard SMF receiver. The numerical and experimental evaluation of the SMF receiver with a CW signal shows that signal fading in unavoidable (up to 30 dB in simulation). The signal fading limits drastically the BER for modulated signals and generate synchronization loss between the transmitter and the receiver. On the contrary, the MM receiver is much more robust to atmospheric turbulence, and the corresponding resulting communication performances shows much less power penalty at a given BER level, even in strong turbulence regime. Compatibility with WDM operation and both amplitude and phase modulation formats were also demonstrated. The evaluation showed that the number of modes considered for the multimode receiver is a critical aspect and should be chosen carefully for a given optical communication use case to optimize the MM receiver. This evaluation is a first validation of a multimode receiver for FSO communication, and paves the way towards efficient high capacity optical feeder link systems.